\def \SAIT #1 #2 {{\em Mem.\ Soc.\ Astron.\ It.\/} {\bf #1}, #2}
\def \MESS #1 #2 {{\em The Messenger\/} {\bf #1}, #2}
\def \ASTRNACH #1 #2 {{\em Astron. Nach.\/} {\bf #1}, #2}
\def \AAP #1 #2 {{\em Astron. Astrophys.\/} {\bf #1}, #2}
\def \AAL #1 #2 {{\em Astron. Astrophys. Lett.\/} {\bf #1}, L#2}
\def \AAR #1 #2 {{\em Astron. Astrophys. Rev.\/} {\bf #1}, #2}
\def \AAS #1 #2 {{\em Astron. Astrophys. Suppl. Ser.\/} {\bf #1}, #2}
\def \AJ #1 #2 {{\em Astron. J.\/} {\bf #1}, #2}
\def \ANNREV #1 #2 {{\em Ann. Rev. Astron. Astrophys.\/} {\bf #1}, #2}
\def \APJ #1 #2 {{\em Astrophys. J.\/} {\bf #1}, #2}
\def \APJL #1 #2 {{\em Astrophys. J. Lett.\/} {\bf #1}, L#2}
\def \APJS #1 #2 {{\em Astrophys. J. Suppl.\/} {\bf #1}, #2}
\def \APSS #1 #2 {{\em Astrophys. Space Sci.\/} {\bf #1}, #2}
\def \ASR #1 #2 {{\em Adv. Space Res.\/} {\bf #1}, #2}
\def \BAIC #1 #2 {{\em Bull. Astron. Inst. Czechosl.\/} {\bf #1}, #2}
\def \JSQRT #1 #2 {{\em J. Quant. Spectrosc. Radiat. Transfer\/} {\bf #1}, #2}
\def \MN #1 #2 {{\em Mon. Not. R. Astr. Soc.\/} {\bf #1}, #2}
\def \MEM #1 #2 {{\em Mem. R. Astr. Soc.\/} {\bf #1}, #2}
\def \PLR #1 #2 {{\em Phys. Lett. Rev.\/} {\bf #1}, #2}
\def \PASJ #1 #2 {{\em Publ. Astron. Soc. Japan\/} {\bf #1}, #2}
\def \PASP #1 #2 {{\em Publ. Astr. Soc. Pacific\/} {\bf #1}, #2}
\def \NAT #1 #2 {{\em Nature\/} {\bf #1}, #2}

\documentstyle{memsait}
\input epsf.sty
\begin{opening}
\title{EVOLUTION OF TYPE IA SUPERNOVAE ON COSMOLOGICAL TIME SCALES}
\author{Inma Dom\'\i nguez$^1$, Peter H\"oflich$^2$, Oscar Straniero$^3$, 
Craig Wheeler$^2$}
\institute{$^1$University of Granada, Granada, Spain\\
$^2$University of Texas, Austin, USA \\ 
$^3$Osservatorio Astronomico di Collurania, Teramo, Italy}
\date{} 
\end{opening}

\begin{document}

\oddpagefooter{}{}{} 
\evenpagefooter{}{}{} 
\ 
\bigskip

\begin{abstract}
Due to their high luminosity at maximum and degree of homogeneity, Type Ia supernovae 
have been extensively  used for cosmological purpouses, in particular   
 to estimate extragalactic distances and 
the Hubble constant.     
Recently the number of Type Ia supernovae detected at high redshift  has increased,  
 opening the  possibility of determining the  
 mass density parameter, the cosmological  
constant and the deceleration parameter.  The observed supernovae appear to be further than expected -even for 
an empty Universe-, implying 
a low density  Universe and moreover an accelerating  Universe. Among the 
  various uncertainties, we address  the possibility  that 
old supernovae are not equal to current supernovae. From first principles, 
 an evolution of  progenitors with  time is expected. Additionally, some 
observations  show a dependence of the observed properties on  galaxy type 
 and colour. Our aim in this work is to study the outcome of exploding CO white dwarfs following the evolution of the progenitor intermediate mass stars 
 with different masses and metallicities. Once this influence of the progentitor has been  determined, the observations may be corrected. At the present stage of this  
project we are not  able to quantify  this effect properly. 
One result is clear, that the differences at maximum are expected to be small ($\sim$0.2 mag) but this is of the same 
order as  all the evidence for a positive cosmological constant ($\sim$0.25 mag).

\end{abstract}

\section{Introduction}
Type Ia supernovae are among  the best candidates for cosmological 
 research: they reach  a luminosity of the same order as that  
 of their host galaxies, a million times greater than Cepheids.  
Moreover, they are easier to understand  than galaxies, because they are the explosion of a single object. For galaxies, it has been  found that the variations with redshift are mainly due to their own 
 evolution  and not to cosmological effects (Tinsley, 1972; Oke, Gunn and 
 Hoessel 1996).
 Concerning the used of SNIa as standard candles, the  homogeneity of the SN 
 event, or at least of their maximum luminosity at maximum, is a fundamental 
 requirement. In this context, a relation between the maximum luminosity and 
 another  observable property is very usefull.  

Supernovae are not equal; the variation of the luminosity at maximum is 
$\sim$2 mag in B and $\sim$1 mag in V, which is probably an indication 
 of a different progenitor population. It has been shown that just excluding 
 the redder events, by  selecting  
 those  with B-V at maximum smaller than 0.5, reduces the dispersion in the Hubble diagram to values around $\sigma \sim$0.25 (Hamuy et al. 1996b; Riess 
 et al. 1996). Recently this dispersion 
   was 
 further  reduced to $\sigma \sim$0.11,  using 
  the maximum-decline relation 
 (decline rate from maximum during the first 15 days in the B band) and the multicolour light-curve shape (MLCS) 
 methods to determine the maximum luminosity (see Phillips et al., 1993; Hamuy et al., 1996a and Riess et al., 1996).
  These methods provide a local calibration of the absolute luminosity,  which is  used  at high redshift to infer the maximum luminosity.  
 To do this, the SNe at high z are assumed to be equal to the local ones. The validity of 
 this assumption is one of our concerns.
       
The use of the {\it maximum luminosity decline relation} to estimate the 
 maximum luminosity of {\it normal} SNIa at z$\leq$
0.1  and the determination of the distance to  host galaxies, using Cepheids observed by  Hipparcos, have led to  
the successful determination of H$_o$. In addition,  
 the linearity of the Hubble flow has been confirmed. At present, only 
  small discrepancies related 
 to the absolute calibration remain, a conservative estimate gives  H$_o$=60$\pm$10 km/s/Mpc, in which the lower values, H$_o$=56$\pm$5 km/s/Mpc, are 
  {\it preferred} 
 (see Branch 1998 and references therein). 

 These discrepancies in the absolute distances do not affect the use of the 
 high redshift supernovae for the determination of the  cosmological 
 parameters, because in this case just  
relative luminosity distances are employed.
  
The high redshift supernovae are further than expected, even for an empty 
 Universe (Filippenko and Riess 1998; Schmidt et al. 1998). The luminosity distance depends on the different cosmological parameters ($\Omega_{m}$, $\Omega_{\lambda}$ and  $\Omega_{k}$) in a different way. Therefore with sufficient observed supernovae at large redshifts it is possible to determine these parameters 
 individually.  
    The fluctuations in the Cosmic Microwave Background radiation (CMB) give 
limits in the $\Omega_{m}$-$\Omega_{\lambda}$ plane orthogonal to those given by 
the high redshift supernovae. A very important result is that the CMB observations exclude the only value of $\Omega_{m}$ that is compatible with $\Omega_{\lambda}$=0 in agreement with  the observations of high redshift supernovae (Hancock et al. 1998).      
These results are consistent with those obtained from 
 the observations of the Mass/Luminosity ratio of galaxies; to date  1150  objects  have been observed 
 up to z=0.6, which give $\Omega_{m}$=0.19$\pm$0.06 (Carlberg et al. 1997).
Other methods, like the variation of the number density of galactic clusters  with redshift,  are consistent with a value of  $\Omega_{m}$=0.3--0.34 (Bahcall 
 et al. 1997). Taking  into account all these determinations, $\Omega_{\lambda}$=0.62$\pm$0.16 and $\Omega_{m}$=0.24$\pm$0.10 (Lineweaver, 1998).     
The observed number of gravitational lenses required $\Omega_{\lambda}\le$ 0.7 (Falco et al. 1998) for a flat Universe, which is consistent with the proposed values and should not be overlooked. 
 The formation of large scale structures also depends  on $\Omega_{m}$. The {\it sponge like} observed 
 structures would  be already formed at  z = 2, if $\Omega_{m}$ were $\leq$ 0.3, while for  $\Omega_{m}$=1 the formation would  continue at the present time. 
 From the high resolution spectra of distant quasars it is possible to infer the  large scale structures, fitting the  
Lyman-$\alpha$ forest. In this case, one of the possibilities is  a closed  
 Universe  with  just pure baryonic matter and  high 
 $\Omega_{\lambda}$, 1.07. Other possibilities agree with the above mentioned 
 values for $\Omega_{\lambda}$ and   $\Omega_{m}$ (van de Bruck and Priest 1998). 
In summary, all the above mentioned results favour a low density Universe with 
a positive cosmological constant. 

\subsection{Observed population age dependence}
There exits  observational evidence suggesting a 
 dependence of the 
 observed properties of SNIa on  galaxy type and colour (see Hamuy et al. 1996a, 
 Branch et al. 1996; Capellaro et al. 1997; Wang et al. 1997).  
 The SNe rate in spirals is double than in ellipticals and in spirals the SNe rate in the bulge  
 decreases. The observed     
scatter in the SNIa light curves in ellipticals is smaller than in spirals and  the SNe  observed in 
ellipticals are less luminous in average than  
those observed in spirals ($\sim 0.3\pm0.1$ mag in B and V), with 
 magnitudes  similar to  
those occurring  in the outer parts of spirals. The dispersion in the decay 
 rate of the light curve  and 
in the expansion velocities is greater in spirals.  

This may indicate a variety in the progenitor population, which 
would  probably be found on  changing the redshift. If this is the case,  the use of the 
 luminosity decline relation or the MLCS method obtained for local SNe 
   may lead to wrong results at high z. However, these local calibrations  
 are made without taking  into account properties of the host galaxy or of 
 the site within 
 the galaxy. This may imply that 
differences due to different progenitor populations have  already been taken into account in these calibrations.   

\subsection{Evolutionary Effect}
More massive progenitors evolve faster, which affects  both  the stars in the 
progenitor binary system and the binary parameters (distances, radii, conditions for mass overflow and the like). 
A similar situation occurs for lower metallicities, whereby low metal stars evolve 
faster and moreover have smaller radii. 
In the case of  high redshift, we expect a major contribution of massive 
 and low metallicity stars. This would change the characteristics of the
  binary systems  
and the relative contribution of the  different scenarios  to the SNIa population.  

An additional consideration is that  the properties of the interstellar medium are also expected 
 to change. For example, 
current  main contributors such as  low mass red giants would not be present 
 at large 
redshifts.  
The extinction law in the LMC and SMC is  significantly different from 
 that of the Milky Way (Bouchet et al. 1985). Moreover, the extinction law 
  also depends  on the redshift of the absorbing dust cloud (H\"oflich 
 and Khokhlov, 1996). Therefore, if the extinction to local or far SNIa is significant, 
 its dependence on metallicity should be taken into account.

\section{Influence of the Initial Composition of the Exploding White Dwarf}

Type Ia supernovae occur  in all types of galaxies,  including  ellipticals,  in which  
star formation ended some Gyr ago. This implies  that a binary system 
 is required as the progenitor of a Type Ia supernova because  the stars 
 currently   evolving in such  
galaxies are low mass stars, which  when evolving as single stars cannot explode.  
  
As it is {\it well} known, the  system that satisfactorily meet all the observational 
constraints for  Type Ia supernovae is  the thermonuclear explosion of an 
accreting CO white dwarf (Hoyle and Fowler 1960; for theoretical details 
 see Wheeler and Harkness (1990), Woosley and Weaver (1994), Bravo et al. (1993  and 1996), H\"oflich and Khokhlov (1996), Nomoto et al. 
 (1997). Three different candidates may reach this 
 situation:   
the explosion of a CO white dwarf that  reaches the Chandrasekhar mass by 
accretion from a MS
or RG star, called the single degenerate Chandrasekhar mass scenario (Nomoto 
 and Sugimoto, 1977; Nomoto, 1982); the explosion 
of a sub-Chandrasekhar mass CO white dwarf that ignites  He explosively in the outer layers, called the 
double detonation sub-Chandrasekhar mass scenario (Nomoto 1980, Limongi and Tornamb\'e 1991; Woosley and Weaver, 1994, 
 H\"oflich and Khokhlov, 1996); the explosion of a CO white 
 dwarf that reaches the Chandrasekhar mass by merging with another CO white 
dwarf, called the double degenerate scenario (Iben and Tutukov, 1984).

None of these scenarios are  completely satisfactory, because  some  
 fundamental problem is encountered in each of them. For the single degenerate Chandrasekhar mass 
it is the problem posed by Cassisi, Iben and Tornamb\'e (1997; see also Tornamb\'e 
 and Piersanti in this volume) of  the impossibility of growing by      
  accretion of  H  to  more than $\sim$1.1$M_\odot$; consequently  the Chandrasekhar mass cannot 
 be attained. Additionally, the lack of H and  He in the spectra and of 
  radio emission means that  the existence of the companion remains unproven. 
 Sub-Chandrasekhar mass models 
 cannot account for the observational properties of the majority 
 of Type Ia supernovae (H\"oflich et al. 1997, Nunget et al. 1997). 
In the case of the double degenerate scenario, the question is that WD binaries with the proper 
masses are not observed at all, although several surveys have been carried out. Moreover, SPH simulations (Benz et al. 1990; Rasio and Shapiro 1995; 
 Mochkovitch et al. 1997) do not produce  a supernova in this way satisfactorily.  
     
\vspace{1cm} 
\centerline{\bf Table 1 - Delayed Detonation Models}

\begin{table}[h]
\hspace{1.5cm} 
\begin{tabular}{|l|c|c|c|c|c|}
\hline
              &$\rho_{tr}$ &Z$^{h}$/Z$_\odot$  & C/O &E$_{kin}$ &M$_{Ni}$\\
              &$ (10^7$ g/cm$^3)$ &   &      & ($10^{51}$ erg/g) & (M$_\odot$)\\
\hline
DD21c          & 2.7    & 1.0   &  1.0    &  1.32  & 0.69   \\
DD23c          & 2.7    & 1.0   &  2/3    &  1.18  & 0.59   \\
DD24c          & 2.7    & 1/3   &  1.0    &  1.32  & 0.70   \\
DD13c          & 3.0    & 1.0   &  1.0    &  1.36  & 0.79   \\
\hline
\end{tabular}
\end{table}

On the other hand, all the above mentioned problems are relative. The question of weather or not the Chandrasekhar mass is achieved remains  
open; high accretion rates have been  employed by several authors to 
 achieve this and  
 the observed counterparts may have been identified  as super soft X ray sources, SSXRS (van den Heuvel et al. 1994,  
 Rappaport et al. 1994a and 1994b).  
 The sub-Chandrasekhar candidates have {\it normal} progenitors, which  
 may explain some events  
(subluminous and blue).  The most intriguing question now is probably  
 why we do not observe them or why they do not occur.
  The double degenerate scenario lacks an observer counterpart:  
 appropriate double white dwarf  systems have not yet been observed, but these    negative  observational results  maybe consistent with the expected numbers (Isern et al. 1997).

The three scenarios may occur  and all of them may 
contribute to some extent to the Type Ia supernovae 
population.  
In the present study we choose the {\it single degenerate M$_{Ch}$} scenario and, for the description of the burning front, the delayed detonation explosion model (a deflagration 
 that turns into a detonation in the outer layers; see Khokhlov 1991).  Detonation, deflagration or  delayed detonation  may occur. We perform 
  our study for the delayed detonation models, as these models  
  give the best fit to the observations.
  H\"oflich and Khokhlov (1996)  satisfactorily explained the observations (light curves and spectra) of 27 Type Ia Supernovae within the delayed detonation (or its variation, pulsation delayed detonation) explosion of Chandrasekhar mass CO white dwarfs. 
For the delayed detonation models two parameters apply to the description of the velocity of the 
 burning front: the ratio of the deflagration velocity to the local  
 speed of sound  and the transition density (the density at which the deflagration 
 turns into a detonation). 

In order to obtain a first estimate of the possible influence of the progenitors 
on the observed properties, 
 H\"oflich, Wheeler and Thielemann (1998) 
 recently studied the outcome (nucleosynthesis, light curves and spectra) 
of a set of CO white dwarfs in which the amount of C/O and the abundances of heavy elements (heavier than Ca) are parametrized (Z$^{h}$) (see Table 1). For the WD, the total mass 
 (1.4$M_\odot$) and central density (9 10$^9$ g/cm$^3$)  are fixed. 
 For the description of the burning front, the ratio of the 
 deflagration velocity to the 
 speed of sound adopted is 0.03 and, except in one case, the transition density is 
2.7 10$^7$ g/cm$^3$.     
 The values adopted 
  are those that best reproduce  the observed 
 properties (light curves and spectra) of the majority of Type Ia supernovae.

This section summerises the results of these calculations, see H\"oflich, Wheeler and Thielemann (1998) for details. 
As shown in Table 1, decreasing the amount of C (the fuel) decreases 
the kinetic energy (column 5) and the amount of radioactive Ni (column 
 6), compare DD21c with DD23c. The 
Ni reduction is similar to the effect of decreasing the transition 
density (compare DD13c with DD21c). In this case, moreover,  the kinetic 
 energy is smaller, but not as much as in the previous case. 
In both,  the chemical elements are more confined in velocity 
space and the region dominated by Si is narrower. This is due to the fact 
that when the kinetic energy is smaller, the outer layers have more time 
to expand, as in the case of a lower transition density. 
When the {\it metallicity} (elements heavier than Ca) is reduced, the changes 
in the amount of Ni and in the kinetic energy are small (compare DD21c with DD24c) and only  a slightly larger amount of Ni is obtained. 

When the amount of C is reduced, the luminosity at maximum is slightly greater ($\sim$0.08 mag). This is due to the 
fact that when  the kinetic energy is smaller, less adiabatic cooling occurs 
  and more of the stored 
energy contributes to the luminosity. On the other hand, the tail luminosity 
is smaller, as in this epoch the luminosity  directly reflects the 
  deposition of  radioactive energy and the  amount of Co is smaller. Therefore  the 
 light curve decline rate after maximun  
is steeper.  This  means that  if the local relation is used to infer the  luminosity at maximum, this luminosity  would be underestimated ($\sim$0.1--0.3 mag), giving  a wrong 
 luminosity distance.

The change in  {\it metallicity} (DD24c) influences the spectra indirectly.  
For lower metallicities, the amount of free electrons per nucleon, Y$_e$, is smaller implying a lower abundance of  $^{54}$Fe, which  is an important source for the opacity in the UV 
 region. 
The spectra is  modified  for $\lambda \leq$ 4000 {\AA}  and  this  influences  
 the  colours  at different  redshifts  when this modification is 
  shifted into different bands. The uncertainties at redshifts z$\geq$0.2 in  
  $\Delta(B-V)$ are of $\sim$0.3. 
 Nonetheless, these modifications  
  affect the regions with velocities of 10000--12000 km/s  that 
 have been subjected to oxygen burning and  are seen  
 around maximum. 
 Later on, the observed light comes from the inner layers which have been 
 subjected  
to explosive burning and therefore Y$_e$ and the abundances of the neutronised 
 elements are determined by the electron captures and 
 do not  depend on the initial composition of the WD.

\section{Evolution of Intermediate Mass Stars: the CO Core}

All the proposed progenitors for Type Ia supernovae  
 rely on the
 explosion 
of a CO white dwarf.
The mass of the accreting CO white dwarf and 
its chemical composition  influence the pre-explosion  evolution  
 and the explosion itself (energy and nucleosynthesis) 
and consequently the observational 
properties.   

CO white dwarfs are initially the degenerate CO cores of intermediate 
mass stars and their mass and chemical composition depend on the evolutionary   history of the progenitor star.  The FRANEC (Frascati RAphson Newton Evolutionary Code) has been used for all the evolutionary models computed in this work.  
 See Straniero, Chieffi and Limongi (1997) and Chieffi, Limongi and Straniero 
 (1998) for a full description of the code.  

More massive stars have more massive 
 CO cores, while higher metallicities (provided Z$\geq  10^{-6}$) mean  less 
 massive cores. When the CO core formed is greater than $\sim$1.1 $M_{\odot}$,
  C is ignited 
 and burns to O and Ne.   
The maximum mass for which C is not ignited,  
the so-called M$_{up}$, depends on the initial composition. This  decreases 
with decreasing metallicity, to  a certain value (Z$\sim$ 10$^{-6}$, see 
 Tornamb\'e and Chieffi, 1986).  
 We obtained  M$_{up}$=8$M_{\odot}$ for Z=0.02 and 6.5$M_{\odot}$ for Z=0.0001; 
  The final CO core mass and composition are determined by  
 the evolutionary history, for which the treatment of turbulent 
 convection, nuclear reaction rates, the equation of state and opacities 
 are crucial (see Dom\'\i nguez et al. 1998).   

Once the central He is exhausted, the star enters the Asymptotic Giant Branch 
 phase (AGB), usually divided into  the early AGB (E-AGB) and the Thermal Pulse 
 AGB phases (TP-AGB). The E-AGB is characterized by  the establishment  of the He 
 burning within a shell outside the contracting CO core. Such an occurrence 
 has two important effects in the evolution of the star: the CO core grows with 
 time and the energy generated by the He shell expands the outer layers 
 and cools the H burning shell, which is spent or nearly spent. At this moment the convective 
 envelope can penetrate inward, pushing protons towards  the vicinity of the He burning shell 
 where, due to the high temperatures, the H is reignited and the  thermal 
 pulses (TP-AGB) begin. During this phase the recursive re-ignition of the H 
 and He burning shells implies a further increase of the CO core mass. 
 However it is believed that the strong AGB mass loss could prevent the occurrence of a large number of thermal pulses. Notice that without this mass loss, the  CO core size could attain the Chandrasekhar mass. 
 However, the observed maximum luminosity of the AGB stars in our galaxy and 
 in the Magallanic clouds (Weidemann 1987; Wood et al. 1992) and the high mass  loss rates observed at the tip of the AGB indicate that the envelope 
 is lost and few pulses have time to occur. Therefore the mass of the CO white dwarf is essentially determined by the value reached  at 
 the end of the E-AGB phase (Vassiliadis and Wood, 1993). 

For the bigger masses (M$\geq$5M$_\odot$ , for Z=0.02 or M$\geq$4M$_\odot$ for Z=0.0001) the H shell is completely extinguished and the inward penetration of the convective envelope enters the He core, which  is called the 2$^{nd}$ dredge up.

\subsection{CO core mass}

If we compare the results of different stellar evolutionary codes,  
differences of up to 50$\%$ in  central He burning life times are found 
(see Dom\'\i nguez et al. 1998). The longer  the central helium burning 
 life time, the greater is the time in which the H shell  advance in mass and 
 the increment of the He core mass will be greater.    
The convective algorithm is important; for example, if we compare the central He burning time in two models, one without overshooting, semiconvection or  
 breathing pulses and the other, like ours, with just semiconvection, the He burning 
 life time is double in the second case. To decide between different mixing 
 hypotheses, the RGB, HB and AGB theoretical life time ratios are compared 
 to the observed stellar number ratios of well studied galactic globular 
 clusters (see Renzini and Fusi Pecci, 1988). The semiconvection scheme 
 satisfactorily reproduces  the observations but a moderate amount of mechanical overshooting cannot be ruled out. 

The current uncertainty about  the rate of the 
 $^{12}C(\alpha,\gamma)^{16}O$ reaction introduces an additional problem. 
The larger  the reaction rate, the longer the 
 He burning life time. More quantitatively, the He burning life time could 
 change by  5--10$\%$.     
  For the bigger masses, a different phenomenon will contribute to 
 determine the final CO core mass. In fact, in such a star, the growth of 
 the CO core at the end of the E-AGB is frozen by the occurrence of 
 the 2$^{nd}$ dredge-up. Again  the treatment of  convective instabilities
  is crucial. 

\vspace{1cm}
\centerline{\bf Table 2 - Properties of the CO core during the TP-AGB phase}

\begin{table}[h]
\hspace{1.5cm} 
\begin{tabular}{|l|c|c|c|c|c|c|c|c|c|}
\hline
M$_T$ &
Z & 
Y & 
M$_{CO}$ & 
 C$_{C}$ &
 C/O$_{core}$ & 
 C/O$_{Mch}$ & 
Rate &
TP (number) \\ 
\hline
 1.5   & 0.001 & 0.23 & 0.560 & 0.260 & 0.533 & 0.781 & CF85 & 3 \\
 3.0   & 0.001 & 0.23 & 0.738 & 0.289 & 0.610 & 0.772 & CF85 & 2 \\
 5.0   & 0.001 & 0.23 & 0.907 & 0.316 & 0.563 & 0.691 & CF85 & 1 \\
 6.0   & 0.001 & 0.23 & 0.976 & 0.320 & 0.537 & 0.650 & CF85 & 1 \\
\hline
 3.0   & 0.0001 & 0.23 & 0.791 & 0.286 & 0.587 & 0.741 & CF85 & 1 \\
 3.0   & 0.001 & 0.28 & 0.815 & 0.297 & 0.623 & 0.760 & CF85 & 8 \\
 3.0   & 0.02 & 0.28 & 0.561 & 0.232 & 0.522 & 0.778 & CF85 & 3 \\
\hline
 3.0   & 0.001 & 0.23 & 0.750 & 0.279 & 0.570 & 0.742 & B-H96 & 6 \\
 3.0   & 0.001 & 0.23 & 0.727 & 0.654 & 2.314 & 1.525 & B-L96 & 5 \\
\hline
 3.0   & 0.001 & 0.23 & 0.759 & 0.289 & 0.621 & 0.773 & CF85 & 11 \\
\hline
\end{tabular}
\end{table}

\subsection{C/O gradient} 

The C/O gradient within the core is {\it built} in successive phases with 
 different characteristics: convective 
central He burning (HB), radiative shell He burning during the early AGB phase 
 and {\it rapid} convective shell He burning during the TP-AGB phase. Once 
 enough He is depleted and C formed, the burning also occurs through the
 $^{12}C(\alpha,\gamma)^{16}O$ reaction. 
  For higher 
 temperatures, the 3$\alpha$ reaction dominates. This is the reason why the central C is greater 
 for more massive stars, in which  less C is destroyed by the $^{12}$C--$\alpha$ capture. Just outside the  convective He burning core,  an instability develops due 
 to the chemical gradient, called  semiconvection. Note that, in some way a small amount 
  of convective core overshoot  mimics the global effect of 
 semiconvection.  Toward the end of the central He burning, when the 
 He mass fraction in the convective core is reduced to 0.1, the 
 occurrence of breathing pulses (Castellani et al. 1985) could delay the 
 He exhaustion. This 
 allows more time for the $^{12}$C--$\alpha$ capture reaction to act and significantly 
 less C is obtained, while the final CO core is slightly bigger. 
 In our code these pulses are inhibited, taking into account 
  observational constraints (Caputo et al. 1989).  
      
The temperature of the He burning shell is higher than that of the central 
 He burning. Nonetheless, more C is left by the shell burning than by the 
 central C burning. This is due to 
 the rapid advance of the shell, as the $^{12}C-\alpha$ does not have much 
 time to convert C into O. As the burning is radiative, a pronounce C gradient 
 is {\it built} during the E-AGB. This gradient is steeper for the smaller masses 
 of our interval. The temperature in the convective burning 
 He shell during the TP phase is higher but these  pulses are very 
 rapid  and the C is not burnt, so that a  C/O$\geq$1 is obtained. 
 However, it is believed, as explained in the introduction, that 
 few pulses occur. Therefore the contribution of the TP phase to the CO core 
 mass and composition is not significant. The carbon abundance gradient 
 within the 
 CO core for the 3M$_\odot$ (Z=0.001; Y=0.23) model is shown in Figure 1. 
 The final average C abundance within the 
 CO core (see column 6, C/O$_{core}$, Table 2) decreases for greater progenitor masses, from 
 3M$_\odot$ to 6M$_\odot$. The maximum 
 difference obtained is of 13$\%$. When metallicity is varied, a maximum difference in the C abundance of 17$\%$ is obtained. In both cases, the differences are 
 smaller than those found in the parametrized models (H\"oflich, Wheeler and  Thielemann, 1998).  

\vspace{1cm}
\begin{figure}
\epsfysize=6cm 
\epsfxsize=8cm 
\hspace{3.5cm}\epsfbox{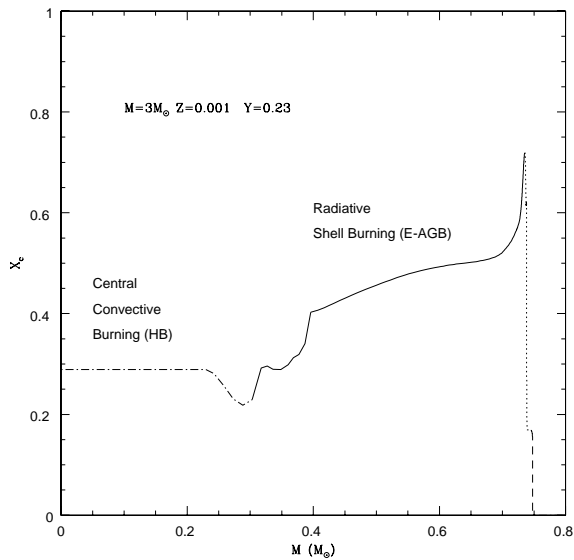} 
\caption[h]{Carbon abundance (mass fraction) at the beginning of the TP-AGB 
 phase. The different evolutionary phases, HB, E-AGB and TP-AGB produce a  
 different carbon abundance gradient.}
\end{figure}

For the accreted matter from the companion, a C/O abundance equal to 1 is 
 assumed (Nomoto et al., 1984). This point is weak and critical, especially for the 
 smaller masses of our interval for which the mass accreted to reach the 
 Chandrasekhar mass  is greater  
  than the initial  CO core mass.  In fact, the average C abundance (see column 7, C/O$_{Mch}$, in Table 2)  
 is finally larger in  the 1.5M$_\odot$ model than in the  3M$_\odot$ model, due to this accreted {\it top}. However, this is probably the correct 
 assumption. In fact, we know from AGB models that the C/O left by the 
 thermally pulsing He  shell,  a situation that 
 resembles the burning of He above the CO white dwarf, is just slightly greater than 1. When the average C abundance in the Chandrasekhar mass WD is considered, the differences between different progenitor masses are more pronounced,  
 18$\%$, as the accreted matter is more significant  for the smaller masses. 
 On the other hand, when the same operation is performed for progenitors 
 with different metallicities, the differences in C abundance tend to be 
 eliminated  as the amount of accreted matter is smaller for the models with lower 
 metallicities for which the amount of C is bigger, until finally between the most 
 extreme cases in metal content, the difference is reduced to 5$\%$.       

In summary, the phase of accretion onto the CO WD should  be simulated  
in a more realistic way, as the chemical profile {\it acquired} during this phase has an 
important weight in the average C abundance.  
 Moreover, the amount of C  will influence not only the conditions at the ignition point but also  
  the propagation of the burning front (see Bravo et al. 1996). It will be 
 very important to clarify wether it is possible  to  
  reach  the Chandrasekhar mass by H accretion.  
 If the answer is negative, the single degenerate Chandrasekhar mass 
 scenario, in spite of all its advantages, should be abandoned and the merging of CO white dwarfs would remain  as the only alternative.  
We are studying the  possibility of  constraining the scenario, taking into 
 account the different C/O expected in the outer layers. For example, the double 
 degenerate scenario would  show a dependence on  metallicity, while 
 this dependence would  not appear in the single degenerate scenario.   
 
It is obvious that the rate adopted  for the  $^{12}C(\alpha,\gamma)^{16}O$ 
 is  very important. By Taking  the high and low rates obtained by Buchmann 
 (1996, 1997), the average C/O abundance within the CO core 
 is 4 times greater for the lower rate (see models B-H96 and B-L96 in Table 2). 
 In the FRANEC code, the adopted rate is that of  Caughlan and  Fowler  
 (1985, CF85 in Table 2) which is  close to the high rate. The correct rate may be 
  somewhere  in between, so our C/O abundances within the CO core could 
 be  lower limits.

\section{Conclusions}
Going back in time, we expect a major contribution of massive  and low 
metallicity stars. 
 This would change the characteristics  of  systems with mass 
overflow (less metallicity means more compact stars with bigger CO cores) 
and the chemical C/O profile within the CO core. From our results, we expect, 
 on average, less C 
for both  more massive and low metal stars. The differences expected 
 are smaller  than  20\%. When this is compared with the results 
of the parametrized models, a difference of $\sim$ 0.2 mag is expected for
 the  
inferred maximum luminosity. This is small but it is of the same order as  
the differences in maximum luminosity that allows us  to discern  the 
different values of the mass density parameter and the cosmological constant 
($\sim $ 0.25 mag). Following our models, the maximum absolute magnitudes  
  of high redshift supernovae are being underestimated, which  would mean that 
 these supernovae are even  
 further.  Nevertheless, it should be remembered that this result is just a first step and important points remain to be  
be clarified. A critical factor  is the composition adopted for the accreted matter. Open  questions are the relations between  progenitor mass 
 and redshift and metallicity  and redshift. The dependence of metallicity 
 on  redshift is 
 highly uncertain.  A  detailed study of the host galaxies is 
 needed but it is the situation when the progenitor was formed that  is 
 most relevant.     

We are currently computing the outcome (explosions, light 
curves and spectra) of the models shown in Table 2. The influence of the C abundance gradient on the propagation of the burning front is an important issue. The amount of fuel, carbon,  directly changes the energy  of the explosion (both kinetic and radioactive energy) although a different  parametrization of the burning 
 front would have the same consequences. Metallicity may have an 
 important effect  on the explosion,  although 1D explosion models limit  our possibilities of taking this into 
 account properly.

 On examining the  pre-supernova and 
   supernova evolution, several  uncertainties remain and for  cosmology 
 studies a difference of just 0.2 mag is significant.
 
 The observations at high redshift show a very homogeneous event.  {\it Why are  
 they so homogeneous (or apparently so) when there are  so many theoretical
  ways for them to vary ??}

\end{document}